\documentstyle[12pt]{article}
\def\be{\begin{equation}}
\def\ee{\end{equation}}
\def\eps{\epsilon}
\def\a{\alpha}
\def\l{\lambda}

\def\ket#1{|{#1}\rangle}  
\def\bra#1{\langle{#1}|}  
\def\norm#1#2{\langle{#1}|{#2}\rangle}   
\def\gov#1#2#3#4{\pmatrix{{#1}&{#2}\cr{#3}&{#4}}}

\begin{document}
\begin{titlepage}
\rightline{IMSc/93/50}
\rightline{\em Dec 93}
\baselineskip=24pt
\begin{center}
{\LARGE Knot invariants from rational conformal field theories}

\vspace{.5cm}

{\bf P. Ramadevi, T.R. Govindarajan and R.K. Kaul} \\
The Institute of Mathematical Sciences, \\
C.I.T.Campus, Taramani, \\
Madras-600 113, INDIA.
\end{center}

\noindent {\bf Abstract}

A framework for studying knot and link invariants from any rational
conformal field theory is developed. In particular, minimal
models, superconformal models and $W_N$ models are studied. The invariants
are related to the invariants obtained from the Wess-Zumino models
associated with the coset representations of these models. Possible
Chern-Simons representation of these models is also indicated. This
generalises the earlier work on knot and link invariants from
Chern-Simons theories.
\vfill
\hrule
\vskip1mm
\noindent{\em email:~rama,trg,kaul@imsc.ernet.in}
\vskip1cm
\end{titlepage}

{\bf 1. Introduction}

In the last few years there has been enormous interest in
obtaining the invariants for knots and links embedded in 3 manifolds
through Chern Simons theories\cite{sch} - \cite{etc}.
It was Schwarz who first conjectured that Jones polynomials\cite{jon} may be
related to Chern-Simons theories\cite{sch}. Witten in his original
papers set up a
general framework to study knots and links through Chern-Simons theories.
He has shown that Jones polynomials are given by the expectation values
of the Wilson loop operators carrying spin $1/2$ representation in
three-dimensional $SU(2)$ Chern-Simons theory. The two-variable
generalization\cite{oce} of Jones invariants are obtained in an
$SU(N)$ Chern-Simons theory when N-dimensional representations are
placed on all the component knots\cite{wit}. These results were
demonstrated by proving that these invariants satisfy the same
Alexander-Conway skein relations as the respective polynomials.
 Unfortunately the
generalized skein relations cannot be recursively solved
for arbitrary knots/links when other than defining representations
are living on the strands. For this purpose
general methods of obtaining these link invariants directly have been
developed. One such method for links made up of upto four strands in $SU(2)$
Chern-Simons theory has been presented in \cite{ours}.
Generalization to arbitrary compact semi-simple groups, in particular
$SU(N)$ has also been obtained\cite{ours1}. A complete solution for
links made up of arbitrary strands has also been developed in
\cite{kaul}. Intimate relation between Chern-Simons theories and
two-dimensional Wess-Zumino conformal field theory plays an important
role in these studies. The Chern-Simons functional integral over
3-manifold with boundary are expressed in terms of vectors in the
Hilbert space of the conformal blocks of the
associated Wess-Zumino conformal field theories on the boundary.
Since primary fields of the Wess-Zumino conformal fields are in one to
one correspondence with the irreducible representation of quantum
groups, there is also a close relationship of Chern-Simons theories with
the quantum groups. The deformation parameter $q$ of these quantum groups
is related to the coupling constant $k$ of our Chern-Simons theories
through $q(k)=\exp {2\pi i \over k+C_v }$, where $C_v$ is the quadratic
casimir of the adjoint representation. Just like Chern-Simons (and
equivalently Wess-Zumino conformal field theories) yield link
invariants, there are  invariants also associated with other rational
conformal field
theories. This can be done by using the representations of braid
generators and an explicit definition of trace with Markov invariances
as presented in ref.\cite{kaul2} for such models. Isidro, Labastida
and Ramallo also
have studied invariants for toral knots from the minimal models
in ref.\cite{lab}. The braiding matrices have been shown to be
factorised in terms of $SU(2)$ Wess-Zumino model braid matrices. The
method of obtaining link invariants presented in
refs.\cite{ours,ours1,kaul} can readily be generalised to any rational
conformal field theory. This will allow us to obtain invariants not only
for toral knots but for any arbitrary multicoloured link. In this paper,
 we shall present the results for minimal, superconformal and $W_N$
theories. The invariants for any arbitrary link in these models will be
shown to be expressible in a simple manner in terms of the invariants
of Wess-Zumino models associated with the coset representation of these
conformal field theories. The underlying quantum group structure will
also be emphasised.

In sec.2 we shall take up the minimal conformal field theories. The
braiding matrices as well as duality matrices will be shown to be
factorised in terms of those of Wess-Zumino models in the coset
representation $(SU(2)_k \otimes SU(2)_1) /SU(2)_{k+1}$ of these
models. The link invariants will also be shown to factorise in the same
manner. In
sec.3, we present a brief discussion of these ideas for the superconformal and
$W_N$ models through their coset representation $(SU(2)_k \otimes
SU(2)_2)/SU(2)_{k+2}$ and
$(SU(N)_k \otimes
SU(N)_1)/SU(N)_{k+1}$ respectively.
In sec.4, we give the possible Chern-Simons description of these general
$(G_k\otimes G_l)/ G_{(k+l)}$ coset models.
\vskip1cm
\noindent{\bf 2. Link invariants from minimal models}

The minimal model fields admit a coset
construction \cite {gko,rav} based on \\
$G/H=(SU(2)_k\otimes SU(2)_1)/SU(2)_{k+1}$.
The central charge {\em c} is given for these models by
$
c = 1 - {6\over (k+2)(k+3)}
$.
The primary fields $\Phi _{(r,s)}$
are labelled by two integers $r$ and $s$ and their conformal
weights are given by
\be
h_{r,s} = {[(k+3)r-(k+2)s]^2 - 1\over 4(k+2)(k+3)}, ~~ 1\leq r\leq
k+1,~ 1\leq s\leq k+2
\ee
There is a doubling of fields through this labelling which can be
distinguished through the condition whether $r+s$ is even or odd.
It is convenient for our future discussion to rewrite these conformal
weights as:
\be
h_{r,s} = h_r^{(k)} - h_s^{(k+1)} + {(r-s)^2\over 4} \label {conf}
\ee
where $h_r^{(k)}$ and $h_s^{(k+1)}$ are the conformal weights of $r$ and
$s$ dimensional
representations of $SU(2)_k$ and $SU(2)_{k+1}$ WZW models respectively
and
$
h_r^{(k)} = {r^2 - 1\over 4(k+2)}
$.
 An explicit labelling of the primary fields which also involves
$SU(2)_1$ description, is done with the help of the label $\eps~~(\eps
=1,2)$ corresponding to the primary fields of $SU(2)_1$.
Incorporating the conformal weight of this level 1 field $(\eps )$ in
the conformal
dimension of the minimal model, we write (\ref {conf}) as:
\be
h_{r,s} = h_r^{(k)}+h_{\epsilon }^{(1)}-h_s^{(k+1)}+ {(r-s)^2 -
(\eps - 1)^2\over 4}    \label {b}
\ee
where $\eps = 1$ for $r+s$ even and $\eps = 2$ for $r+s$ odd.

The fusion rules satisfied by these primary fields is
\be
\Phi_{(r_1,s_1)} \otimes \Phi_{(r_2,s_2)} = \sum_{r_3=|r_1-r_2|+1}^{min(r_1+r_2
-1,2k+3-r_1-r_2)}~~ \sum_{s_3=|s_1-s_2|+1}^{min(s_1+s_2
-1,2k+5-s_1-s_2)}\Phi_{(r_3,s_3)} \label{a}
\ee
The modular transformation matrix for the minimal model is given by
\be
S_{rs}^{r's'}~~=~~\sqrt{{8\over (k+2)(k+3)}}~(-1)^{(r+s)(r'+s')}sin~{\pi
rr'\over k+2}~sin~{\pi ss'\over k+3}
\ee
The modular transformation is related to the unknot polynomials for any
conformal field theory\cite{etc,alva}. This unknot invariant
associated with representation $\Phi _{(r,s)}$ is given by
$V_{(r,s)}[U]~=~S^{11}_{rs}/S^{11}_{11}$. The factorised form of this
modular transformation above imply
\be
V_{(r,s)}[U]~~=~~V_r^{(k)}[U]~ V_s^{(k+1)}[U]
\ee
where the two unknot invariants on the right hand side correspond
to r and s dimensional representations of $SU(2)_k$ and $SU(2)_{k+1}$
Wess-Zumino models respectively.
Notice that
$V_\eps ^{(1)}[U]~=~1$ for both $\eps =1$ and 2.
This factorised form of the unknot invariant is consistent with the fact
that link invariant for two cabled knots (such as unknots here) obey the
fusion rules(\ref{a}) of the theory.
\be
V_{(r_1,s_1)}[
U]V_{(r_2,s_2)}[U]~=\sum
_{r_3,s_3}
V_{(r_3,s_3)}[U] \label {c}
\ee

To study the link invariants, we need the braiding properties of the
correlator-conformal blocks.
The eigenvalues of the braiding matrix $B[(r_1,s_1);(r_2,s_2)]$ as depicted
below are
given by
\setlength{\unitlength}{2cm}
$$
\begin{picture}(8,2)
\put(1,0){\line(0,1){.9}}
\put(1,.9){\line(1,1){.55}}
\put(.45,1.45){\line(1,-1){.55}}
\put(1.8,.4){\vector(1,0){1.1}}
\put(3.5,0){\line(0,1){.5}}
\put(3.5,.5){\line(1,1){.3}}
\put(3.2,.8){\line(1,-1){.3}}
\put(3.2,.8){\line(1,1){.6}}
\put(3.55,1.05){\line(1,-1){.25}}
\put(3.2,1.4){\line(1,-1){.25}}
\put(6.2,0){\line(0,1){.9}}
\put(6.2,.9){\line(1,1){.55}}
\put(5.65,1.45){\line(1,-1){.55}}
\put(.2,1.55){$\Phi _{(r_1,s_1)}$}
\put(1.35,1.55){$\Phi _{(r_2,s_2)}$}
\put(1.5,.6){$B[(r_1,s_1);(r_2,s_2)]$}
\put(2.9,1.55){$\Phi _{(r_2,s_2)}$}
\put(3.6,1.55){$\Phi _{(r_1,s_1)}$}
\put(3.8,.4){$=$}
\put(4.1,.4){$\lambda_{(r_3,s_3)}[(r_1,s_1);(r_2,s_2)]$}
\put(5.4,1.55){$\Phi _{(r_2,s_2)}$}
\put(6.55,1.55){$\Phi_{(r_1,s_1)}$}
\put(1.05,0){$\Phi _{(r_3,s_3)}$}
\put(3.55,0){$\Phi _{(r_3,s_3)}$}
\put(6.25,0){$\Phi_{(r_3,s_3)}$}
\end{picture}
$$
\be
\lambda _{(r_3,s_3)}[(r_1,s_1);(r_2,s_2)]~~=~~\Omega~\exp {i\pi
(h_{r_1,s_1}+h_{r_2,s_2}-h_{r_3,s_3})}
\ee
where $\Omega ~=~\pm 1$ and we choose it to be
$(-1)^{f_1+f_2-f_3}~~$ where
$f_i$ is an integer given in terms of the dimensions of the representations as
$f_i ={1\over 2}(r_i-s_i+\eps_i -1)({r_i-s_i-\epsilon_i+1\over
2}+1)$.

Using eqn (\ref {b}), these  eigenvalues  can be rewritten in
explicitly factorized form in  terms of the WZW model braiding eigen
values as shown:
\be
\lambda_{(r_3,s_3)}[(r_1,s_1);(r_2,s_2)]~~=~~
\lambda _{r_3}^{(k)}(r_1,r_2)~~(\lambda
_{s_3}^{k+1}(s_1,s_2))^{(-1)}~~\lambda_{\epsilon _3}^{(1)}(\epsilon
_1,\epsilon _2)
\ee
where $\l _{r_3}(r_1,r_2)=(-1)^{{1\over 2}[(r_1-1)+(r_2-1)-(r_3-1)]}
\exp i\pi (h_{r_1}^{(k)}+h_{r_2}^{(k)}-h_{r_3}^{(k)})
$ and $\eps _i$ are the $SU(2)_1$ labels of the primary fields $\Phi
_{(r_i,s_i)}$ respectively.
The knots and links have  to be specified with some framing convention.
 Usual
convention corresponds to standard framing where linking number of the
knot and its frame is zero. Twisting strands does not leave this framing
unaltered.
However, we shall here use the vertical framing \cite {kauf} where braiding
does not alter the
framing. In this framing, link invariants reflect only regular isotopy
invariance instead of invariance under all the Reidmeister moves. This is
reflected in the following equation for the $SU(2)_k$ fields:
$$
\begin{picture}(8,1)
\put(2.5,0){\line(1,1){.3}}
\put(2.2,.3){\line(1,-1){.3}}
\put(2.2,.3){\vector(1,1){.6}}
\put(2.55,.55){\line(1,-1){.25}}
\put(2.2,.9){\line(1,-1){.25}}
\put(3.5,.5){$=~~(-1)^r~e^{-i\pi h_r}$}
\put(5,.5){\vector(1,0){1.5}}
\put(5.7,.2){$r$}
\end{picture}
$$
The corresponding equation for minimal models involves the conformal weights
($h_{r,s}$) of
the primary fields of these models:
$$
\begin{picture}(8,1)
\put(2.5,0){\line(1,1){.3}}
\put(2.2,.3){\line(1,-1){.3}}
\put(2.2,.3){\vector(1,1){.6}}
\put(2.55,.55){\line(1,-1){.25}}
\put(2.2,.9){\line(1,-1){.25}}
\put(3.5,.5){$=~~(-1)^{2f_1}~e^{-i\pi h_{r_1,s_1}}$}
\put(5.5,.5){\vector(1,0){1.5}}
\put(5.9,.2){$\Phi _{(r_1,s_1)}$}
\end{picture}
$$
where $h_{r,s}$ has been written in terms of eqn(\ref {b}).

Besides framing, we shall also put orientation on the lines in a link.
 Any link can be thought of as closure of oriented braids. For
this purpose we need to distiguish braiding in parallel and anti
parellel strands. The eigenvalues for these braiding matrices are
related. For example for $SU(2)_k$ theory, the eigenvalues
of a right handed half twist in parallel (+) and antiparallel (-) strands
are given for standard framing as in ref(\cite {ours,ours1,kaul}).
In vertical
framing these are simply related; they are each others inverse:
\be
\l^{(+)}_{r_3}(r_1,r_2)=
\l_{r_3}(r_1,r_2)~=~
[\l^{(-)}_{r_3}(r_1,r_2)]^{-1}.
\ee
Same is true for the eigenvalues of braiding matrices in minimal
models.
\be
\l^{(+)}_{(r_3,s_3)}[(r_1,s_1);(r_2,s_2)]=
\l_{(r_3,s_3)}[(r_1,s_1);(r_2,s_2)]=
[\l^{(-)}_{r_3,s_3}(r_1,s_1;r_2,s_2)]^{-1}
\ee

Factorisability of braid matrices \cite {blok,moor,lash}
can also be directly justified from
general properties of link invariants. This we shall discuss later.
The method of obtaining link invariants for $SU(2)$ and $SU(N)$ Chern-
Simons theories presented in ref.(\cite{ours,ours1})
can be immediately generalised
to any rational conformal field theory. We shall now present
a brief discussion
of this formalism in the context of minimal models.

Consider a 3-ball having 4-punctures with two strands
going into the boundary $(S^2)$ carrying primary fields $\Phi_{(r_1,s_1)}$
and $\Phi _{(r_2,s_2)}$ of the minimal models and the two strands
coming out of the boundary carrying the appropriate primary fields with
no crossing as shown in the fig.1.
With this diagram, we associate a state $\ket {\psi_0}$
in the Hilbert space of conformal blocks of four-point correlators
on $S^2$. This state can be
written in a convenient basis. We express it in terms of the
eigen-basis
of braid operators twisting central two strands or side two strands as:
\be
\ket {\psi_0}=\sum _{r_3,s_3}
\mu_{(r_3,s_3)}~\ket {\phi _{(r_3,s_3)}^{cent}}
{}~=~\sqrt {[r_1][r_2][s_1][s_2]}~~\ket {\phi _{(1,1)}^{side}} \label {one}
\ee
The summation over $r_3$ and $s_3$ runs over values allowed by fusion
rules eqn.(\ref {a}). Here square brackets denote
 q-number defined as $[n]=(q^{n/2}-q^{-n/2})/(q^{1/2}-q^
 {-1/2})$.
The corresponding state in the dual Hilbert-space representing
the same fig.1
with oppositely oriented boundary is given by
\be
\bra {\psi_0}=
\sum _{r_3,s_3}
\mu_{(r_3,s_3)}~\bra {\phi _{(r_3,s_3)}^{cent}}
{}~=~\sqrt {[r_1][r_2][s_1][s_2]}~~\bra {\phi _{(1,1)}^{side}} \label {two}
\ee
Glueing the diagrams of
fig.1 
onto its dual along the boundary gives us
two disjoint unknots. This is represented by the inner product of the
assoiciated states as:
\be
V_{(r_1,s_1)}[U]~~
V_{(r_2,s_2)}[U]~~
=~~\norm{\psi_0}{\psi_0}~~=~~ \sum _{r_3} \sum _{s_3}\mu _{(r_3,s_3)}^2
\ee
In view of eqn (\ref {c}), this implies
$\mu _{(r_3,s_3)}~=~\mu _{r_3}~\mu _{s_3}$ where $\mu _{r_3}$ and
$\mu_{s_3}$ are the
corresponding coefficients for the $SU(2)_k$ and $SU(2)_{k+1}$ WZW models: $\mu
_r= \sqrt{[2r+1]}$. In order to appreciate factorisation of braid
 matrices, consider
the state $\ket
{\psi _m}$ drawn in fig.2,  representing $m$ half-twists in central two
strands in parallel orientation. This state expanded in terms of the above
basis can be written
as:
\be
\ket {\psi_m}=(B)^m\ket {\psi _0} =\sum _{r_3,s_3}
\mu
_{(r_3,s_3)}(\l^{(+)} _{(r_3,s_3)}[(r_1,s_1);(r_2,s_2)])^{-m}
{}~\ket {\phi _{(r_3,s_3)}^{cent}}
\ee
where $\l^{(+)} _{(r_3,s_3)}[(r_1,s_1);(r_2,s_2)]$ are the eigenvalues of the
braid matrix for parallely oriented strands.
Similarly, we can associate a state $\ket {\psi _0^{\prime}}$ with the
fig.3 and
writing in terms of eigen basis of braid operator twisting side two
strands or central two strands we get:
\be
\ket {\psi^{\prime }_0 }=\sum _{r_3,s_3}
\mu
_{(r_3,s_3)}
{}~\ket {\phi _{(r_3,s_3)}^{side}}
{}~=~\sqrt {[r_1][r_2][s_1][s_2]}~~\ket {\phi _{(1,1)}^{cent}} \label
{thre}
\ee
We can  write $\ket {\psi _m ^{\prime }}$ representing $m$-half twists
in
the side two antiparallel strands as shown in fig.4 as
\be
\ket {\psi^{\prime } _m}=(B)^m\ket {\psi _0} =\sum _{r_3,s_3}
\mu
_{(r_3,s_3)}(\l^{(-)} _{(r_3,s_3)}[(r_1,s_1);(r_2,s_2)])^m
\ket {\phi _{(r_3,s_3)}^{side}}  \label {four}
\ee

Now look at the closure of two-strand braid carrying same
primary field $\Phi _{(r_1,s_1)}$ with one half twist in the central two
strands.
This represents
unknot and is given by
\be
V_{(r_1,s_1)}[U](-1)^{2f_1}\exp(\pm i\pi 2h_{r_1,s_1})
=\norm{\psi _0}{\psi _1}
=\sum _{r_3,s_3}~[2r_3+1][2s_3+1](\l
^{(+)}_{(r_3,s_3)}[(r_1,s_1;r_1,s_1)])^{\pm 1}
\ee
This equation is consistent with factorisation of the braid eigenvalues
in terms of $SU(2)_k$, $SU(2)_{k+1}$, $SU(2)_1$ eigenvalues.

Another interesting consistency condition is obtained by looking at a
Hopf (anti Hopf) link in two equivalent representations as shown in fig.5.
In the first one, we have braiding in parallely oriented two strands
whereas in the second antiparallel strands are braided. The two ways of
writing this invariant have to be equal:
\be
\sum _{r_3,s_3}[2r_3+1][2s_3+1](\l _{(r_3,s_3)}^{(+)}[(r_1,s_1);(r_2,s_2)])^
{\pm 2}
=\sum _{r_3,s_3}[2r_3+1][2s_3+1](\l _{(r_3,s_3)}^{(-)}[(r_1,s_1);(r_2,s_2)])^
{\pm 2}
\ee

There are  two copies of each primary field in the minimal model which needs to
be identified up to a phase:
$\Phi _{(r_1,s_1)}~
=~\exp\{{\pi i \over 4}(r_1-s_1-\eps _1-1)(\eps _1-\eps _2)\}
\Phi _{(k+2-r_1,k+3-s_1)}$ where $\eps _2 = 3 -\eps _1$
and accordingly the braid operator
$B$ operator is transformed to $\tilde B$ such that
\be
\langle \Phi _{(r_1,s_1)}\Phi _{(r_2,s_2)}\Phi _{(r_3,s_3)}|B|\Phi _{(r_2,s_2)}
\Phi _{(r_1,s_1)}\phi _{(r_3,s_3)}\rangle
=\langle \Phi _{(r^\prime _1,s^\prime
_1)}\Phi
 _{(r^\prime_2,s^\prime _2)}\Phi _{(r^\prime _3,s^\prime _3)}|\tilde B |\Phi_
{(r^
 \prime _2,s^\prime _2)} \Phi_
{(r^\prime _1,s^\prime _1)}\Phi_ {(r^\prime _3,s^\prime _3)}\rangle \label
{five}
\ee
where the primed states are related to the unprimed ones by the above
transformation along with the requirement that the respective three
fields satisfy the fusion rules. While the braid matrix $B$ for the
unprimed fields factorise in terms of the corresponding $SU(2)$ braid
matrices, $\tilde B$ also factorise in terms of braid matrices
 associated with primed $SU(2)$ fields.

Not only do braid matrices factorise, so do the duality
matrices\cite{kaul2}.
The duality matrices relate the four-point correlators in different
basis as shown below: .
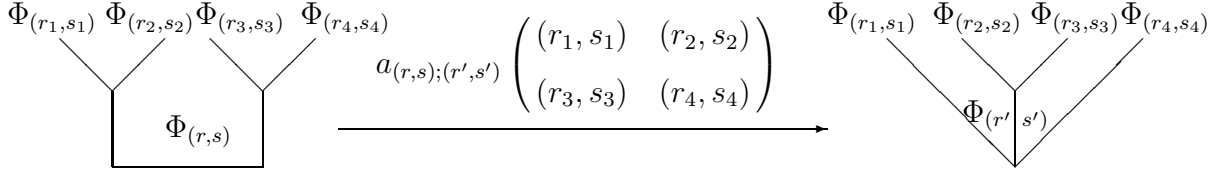
\setlength{\unitlength}{1cm}
$$
\begin{picture}(15,3)
\put(1,0){\line(1,0){2}}
\put(1,0){\line(0,1){1}}
\put(1,1){\line(1,1){.7}}
\put(0.3,1.7){\line(1,-1){.7}}
\put(3,0){\line(0,1){1}}
\put(3,1){\line(1,1){.7}}
\put(2.3,1.7){\line(1,-1){.7}}
\put(4,.5){\vector(1,0){6.5}}
\put(13,0){\line(0,1){1}}
\put(13,1){\line(1,1){.7}}
\put(12.3,1.7){\line(1,-1){.7}}
\put(13,0){\line(1,1){1.7}}
\put(11.3,1.7){\line(1,-1){1.7}}
\put(4.5,1.25){$a_{(r,s);(r',s')}\gov{(r_1,s_1)}{(r_2,s_2)}{(r_3,s_3)}{(r_4,s_4)}$}
\put(-.40,1.9){$\Phi _{(r_1,s_1)}$}
\put(.9,1.9){$\Phi _{(r_2,s_2)}$}
\put(1.7,.4){$\Phi _{(r,s)}$}
\put(3.5,1.9){$\Phi _{(r_4,s_4)}$}
\put(2.1,1.9){$\Phi _{(r_3,s_3)}$}
\put(12.3,.6){$\Phi _{(r'~s')}$}
\put(10.5,1.9){$\Phi _{(r_1,s_1)}$}
\put(11.9,1.9){$\Phi _{(r_2,s_2)}$}
\put(13.2,1.9){$\Phi _{(r_3,s_3)}$}
\put(14.4,1.9){$\Phi _{(r_4,s_4)}$}
\end{picture}
$$
These conformal blocks refer to the bases $\ket {\phi^{side}_{(r,s)}}$
and $\ket {\phi^{cent}_{(r',s')}}$ of eqns.(\ref {one}),(\ref {two}).
Thus,
\be
\norm {\phi _{(r,s)}^{side}} {\phi _{(r',s')}^{cent}}~=~
a_{(r,s);(r',s')}\left (\matrix
{(r_1,s_1) ~~(r_2,s_2)\cr (r_3,s_3)~~
(r_4,s_4)}\right) \label {six}
\ee

The factorisation of these duality matrices in terms of $SU(2)$ duality
matrices can be justified in the following manner. Consider the fig.2
and
the fig.4 with $\em m $ being 1. These have been redrawn in fig.6.
Clearly, the states representing them are equal. Use eqn(\ref {thre}),
(\ref {four}) and (\ref {six}) to write:
$$
(\lambda _{r_3,s_3}^{(+)}[(r_1,s_1);(r_2,s_2)])^{\pm 1}~a_{(r_3,s_3);(1,1)}
\left (\matrix {(r_1,s_1)~~(r_2,s_2)\cr (r_2,s_2)~~(r_1,s_1)}\right) =
{}~~~~~~~~~~~~~~~~~~~~~~~~~~~~~~~~~~~~~~~~~~~~~~~~~~~
$$
\be
{}~~~~~~~~~\sum _{r,s}
a_{(1,1);(r,s)}\left (\matrix
{(r_2,s_2)~~(r_2,s_2)\cr (r_1,s_1)~~(r_1,s_1)}\right)~~(\lambda _{r,s}^{(-)}
[(r_1,s_1);(r_2,s_2)])^{\pm 1}~~a_{(r_3,s_3);(r,s)}\left (\matrix
{(r_2,s_2)~~(r_1,s_1)\cr
(r_2,s_2)~~(r_1,s_1)}\right)
\ee

\noindent This equation is explicitly satisfied by the factorised duality
matrix:
\be
a_{(r,s);(r',s')}\left (\matrix
{(r_1,s_1)~~(r_2,s_2)\cr (r_3,s_3)~~(r_4,s_4)}\right)~=~
a_{r,r'}^{(k)}\left
(\matrix {r_1 ~~r_2\cr r_3~~ r_4}\right)~
a_{s,s'}^{(k+1)}\left
(\matrix {s_1 ~~s_2\cr s_3~~ s_4}\right)
{}~a_{\eps ,\eps '}^{(1)}\left
(\matrix {\eps_1 ~~\eps_2\cr \eps_3~~ \eps_4}\right)~.
\ee

Here
$a_{r,r'}^{(k)}\left
(\matrix {r_1 ~~r_2\cr r_3~~ r_4}\right)$,~
$a_{s,s'}^{(k+1)}\left
(\matrix {s_1 ~~s_2\cr s_3~~ s_4}\right)$ and
$a_{\eps ,\eps '}^{(1)}\left
(\matrix {\eps_1 ~~\eps_2\cr \eps_3 ~~\eps_4}\right)$ are the duality
matrices for $SU(2)_k,SU(2)_{k+1}$ and $SU(2)_1$ theories respectively
and are given in terms of $6-j$ symbols.
\be
a_{r,r'}^{(k)}\left
(\matrix {r_1 ~~r_2\cr r_3 ~~r_4}\right)~=~
(-1)^{{1\over 2}(r_1+r_2+r_3+r_4)}~\sqrt {[r][r']} \left (\matrix
{{r_1-1\over 2} ~{r_2-1\over 2} ~{r-1\over 2}\cr {r_3-1\over 2}
{}~{r_4-1\over 2} ~{r'-1\over 2}}\right ).
\ee

The $6-j$ symbol is given by:
\begin{eqnarray*}
\pmatrix {j_{1} & j_{2} & j_{12} \cr j_{3} & j_{4} & j_{23}}&= &
\Delta(j_1,j_2,j_{12}) \Delta(j_3,j_4,j_{12}) \Delta(j_1,j_4,j_{23})
\Delta(j_3,j_2,j_{23})  \\
&&\sum_{m\geq0}{(-)^{m} [m+1]!}
\Bigl\{ {[m-j_1-j_2-j_{12}]}!\Bigr. \\
&&{[m-j_3-j_4-j_{12}]}!
{[m-j_1-j_4-j_{23}]}!\\
&&{[m-j_3-j_2-j_{23}]}!
{[j_1+j_2+j_3+j_4-m]}! \\
&&
\hskip.5cm\Bigl.{[j_1+j_3+j_{12}+j_{23}-m]}!{[j_2+j_4+j_{12}+j_{23}-m]}!\Bigr\}^{-1}
\hskip1.5cm
\end{eqnarray*}
\noindent and
$$ \Delta(a,b,c) = \sqrt{{{[-a+b+c]![a-b+c]![a+b-c]!}\over{[a+b+c+1]!}}}
$$
\noindent Here $[a]! = [a] [a-1] \ldots [2][1]$.
 The spin triplets
$(j_1,j_2,j_{12})$, $(j_3,j_4,j_{12})$, $(j_2,j_3,j_{23})$ and
$(j_1,j_4,j_{23})$ satisfy the fusion rules of $SU(2)_k$ Wess-Zumino
model.

Since $\Phi _{(r,s)}$ and
$\Phi _{(k+2-r,k+3-s)}$ fields of the minimal models are
identified, the corresponding duality matrices $\tilde a$ for the
shifted fields should also factorise in the same manner as does the
braiding matrix $\tilde B$ of eqn(\ref {six}). This can be easily
verified by using the following properties of the $SU(2)_k$ duality
matrices under the shifts:
$$
a_{rr'}\left (\matrix
{r_1~ r_2\cr r_3~ r_4 }\right )=
(-1)^{{1\over 2}(r_1+r_2+r_3+r_4-2k)}
a_{rr'}\left (\matrix
{(k+2-r_1) ~~(k+2-r_2)\cr (k+2-r_3)~
{}~~(k+2-r_4)}\right )= \hskip1cm
$$
$$
(-1)^{{r_1+r_3-r-r'\over 2}}a_{(k+2-r)(k+2-r')}\left
(\matrix
{(k+2-r_1) ~~~r_2\cr (k+2-r_3)
 ~~~r_4}\right )~~~~= \hskip3cm
$$
$$
\hskip2cm (-1)^{{-r'-1+r_2+r_3\over 2}}a_{(k+2-r) r'}\left (\matrix
{(k+2-r_1) ~~~~~r_2~~~~~~~~~\cr ~~~~~~~~r_3~~~~~ (k+2-r_4)}\right )
$$
\vskip1cm
Factorisation of braid matrices and duality matrices has an
important and immediate implication. This implies that the link invariants
also factorise. We formulate this main result in the form of a theorem as
follows:

\noindent {\bf Theorem 1:}
The invariant for any knot/Link $(L)$ carrying primary fields of the
minimal model $\Phi _{r_1,s_1},\Phi _{r_2,s_2},......,\Phi _{r_n,s_n}
$ on the $n$ component knots is given by:
\be
V_{(r_1,s_1),.......,(r_n,s_n)}[L]~~=~~V_{r_1,.......,r_n}^{(k)}[L]~~
V_{s_1,.......,s_n}^{(k+1)}[\bar L]~~V_{\epsilon_1,.....,\epsilon _n}^
{(1)}[L]~~
\ee
where
$V_{r_1,r_2,...,r_n}^{(k)}[L]~~$ is the invariant
from the $SU(2)_k$ WZW models for the link $L$ carrying  $r_1,r_2,....r_n$
dimensional representations on the component knots.
$V_{s_1,.......,s_n}^{(k+1)}[\bar L]~$ is the corresponding
invariant for $SU(2)_{k+1}$ theory for the mirror link $\bar L$ and
$~V_{\epsilon_1,.....,\epsilon _n}^{(1)}[L]$ is the invariant for the
$SU(2)_1$ model. Here $\eps _i~=~1$ or 2 for $(r_i-s_i)$ even or odd
respectively.

Our discussion above also reflects
the quantum group structure for
the minimal model to be $SU_{q(k)}(2)~\otimes~SU_{q(1)}(2)~
\otimes ~SU_{ q^*(k+1)}(2)$.
This is in agreement with the conjecture for a
general G/H models\cite{blok,moor}. Exploiting the properties of vertex
operators, this quantum group structure has already been pointed out by
Lashkevich \cite{lash}.
\vskip1cm
\noindent{\bf3. Superconformal and $W_N$ models}

Our discussion above is also valid for other conformal field theories.
We shall discuss the $N=1$ superconformal minimal models now.
These models
are obtained through the coset constuction \cite {gko,mussardo} based on
${G/H} = (SU_k(2)\otimes SU_2(2))/SU_{k+2}(2).
$
The central charges for these models are given by
$
c = {3\over 2}\left\{1 - {8\over (k+2)(k+4)}\right\}
$.
The primary fields are again labelled through two integers $r$ and $s$
and the conformal weight of the $\Phi_{(r,s)}$ primary field is given
by
\be
h_{r,s} = {[(k+4)r - (k+2)s]^2 -4\over 8(k+2)(k+4)} + {\a \over 8}
\ee
NS sector is defined through $r-s = 2{\cal Z}$ and $\a = 0$.
Ramond sector is obtained through $r-s = 2{\cal Z}+1$ and $\a =
{1\over 2}$.
The conformal weights can also be rewritten as
\be
h_{r,s} = h_r^{(k)} - h_s^{(k+2)} + {(r-s)^2\over 8} + {\a \over 8}
\ee
Like in the minimal models in Sec.2, we incorporate
the level 2 field labelled
by $\eps $ in the conformal dimensions to get :
\be
h_{r,s} = h_r^{(k)}+h_{\eps}
^{(2)}-h_s^{(k+2)}+{(r-s)^2-(\eps -1)^2 \over 8}
\ee
where for $(r-s)~$ even, $\eps $ is given by
$
\eps = 1 +[(r-s) ~~mod~ 4 ]
$
and for $(r-s)~$ odd, $\eps = 2 $.
The fusion algebra is given by
\be
\Phi_{(r_1,s_1)} \otimes \Phi_{(r_2,s_2)} =
\sum_{r_3=|r_1-r_2|+1}^{min(r_1+r_2
-1,2k+3-r_1-r_2)} ~~~~ \sum_{s_3=|s_1-s_2|+1}^{min(s_1+s_2
-1,2k+7-s_1-s_2)}\Phi_{(r_3,s_3)}
\ee
The modular tranformation matrix for the NS superconformal model is
given by \cite {mussardo,kas}
\be
S_{rs}^{r^{\prime} s^{\prime}}~~=~~{4 \over (\sqrt {(k+2)(k+4))}} \sin {\pi
rr^{\prime} \over (k+2)}~\sin {\pi ss^{\prime}\over (k+4)}
\ee
Like in the minimal models of sec.2 , this
implies that the unknot polynomial $V_{(r,s)}[U]$ for the strands
carrying the superconformal NS field $\Phi _{(r,s)}$ to be
\be
V_{(r,s)}[U]~~=~~V_r^{(k)}[U]~V_s^{(k+2)}[U]
\ee
Note that the level two field in the NS sector are $\eps = 1$ or $3$ and
hence its unknot polynomial ($V_{\epsilon}^{(2)}[U]$) is 1.
For the Ramond field, the unknot polynomial will contain a factor
$V_{\eps}^{(2)}[U]$ with $\eps = 2$. Thus in general:
\be
V_{(r,s)}[U]~~=~~V_r^{(k)}[U]~V_{\eps }^{(2)}[U]V_s^{(k+2)}[U]
\ee

The braiding eigenvalues are given by:
\be
\lambda_{(r_3,s_3)}[(r_1,s_1);(r_2,s_2)]~~=~~\Omega \exp \pi i
(h_{r_1,s_1}+h_{r_2,s_2}-h_{r_3,s_3})
\ee
where $\Omega =\pm 1$ explicitly given by
$
\Omega ~~=~~(-1)^{f_1+f_2-f_3}
$
and $f_i$ are integers and are given in terms of the dimensions of the fields
is
given by
$$
f_i=({r_i-s_i+\epsilon_i -1\over 2})({r_i-s_i-\epsilon_i+1\over
4}+1)
$$
This eigenvalue can be rewritten in terms of WZW model braiding eigen
values as shown:
\be
\lambda_{(r_3,s_3)}[(r_1,s_1);(r_2,s_2)]~~=~~\lambda _{r_3}^{(k)}(r_1,r_2)~~
( \lambda
_{s_3}^{(k+2)}(s_1,s_2))^{-1}~~\lambda_{\eps _3}^{(2)}(\epsilon
_1,\epsilon _2)
\ee

Like in the minimal model in sec.2, we shall use the vertical framing
for the knots. The eigenvalues for braiding matrices associated with
parallel and antiparallel strands are again inverse of each other.
Further duality matrices again factorise in terms of $SU(2)_k$,
$SU(2)_{k+2} $ and $SU(2)_2$ duality matrices. This leads us to the
result:

\noindent {\bf Theorem 2 :}
The invariant for a link L, made up of $n$ knots, carrying
representations
$\Phi
_{(r_1,s_1)},......,\Phi _{(r_n,s_n)}$ respectively is given in terms of
$SU(2)_k $, $SU(2)_{k+2}$ and $SU(2)_2$ invariants as follows:
\be
V_{(r_1,s_1),.......,(r_n,s_n)}[L]~~=~~V_{r_1,.......,r_n}^{(k)}[L]~~
V_{s_1,.......,s_n}^{(k+2)}[\bar L]~~V_{\eps_1,.....,
\eps _n}^{(2)}[L]~~
\ee
where $\bar L$ stands for the mirror link and
$\eps _i =1,3 $ if $r_i - s_i =4{\cal Z}, 4{\cal Z}+2 $ repectively and
$\eps _i=2 $ if $r_i-s_i=2{\cal Z}+1$.

Similar discussion can be extended to $W_N$ models \cite {bilal}, which
have a coset representation in terms of
$SU(N)_k\otimes SU(N)_1/SU(N)_{k+1}
$.
The central charge for these models is given by
$
c~=~(N-1)(1~-~ {N(N+1) \over (k+N)(k+N+1)}).
$
The primary fields are labelled by two weight vectors $\mu, \nu$
of $SU(N)$ and
the conformal weights  for the $\Phi_{(\mu,\nu)}$ field is given
by
\be
h_{\mu , \nu} = {[(k+N+1)n_i - (k+N)m_i+1]g_{ij}[(k+N+1)n_j - (k+N)m_j+1] -
 \rho^2\over 2(k+N)(k+N+1)}~~
\ee
where $\rho $ is the Weyl vector, $g_{ij}$ is the cartan matrix and the
weights in terms of fundamental weights $\Lambda ^i $ of $SU(N)$ are
$\mu = \sum _{i=1}^{N-1} n_i \Lambda ^i$ , $\nu = \sum _{i=1}^{N-1} m_i
\Lambda ^i $. Further $\mu $ and $\nu
$ are restricted to $\mu . \psi \leq k$ ,
$\nu . \psi \leq {k+1}$, where $\psi $ is the longest root.

The primary fields of this model come in N copies, which are identified
upto a phase.
These copies go into each other under ${\cal Z}_N$
transformation. We can easily generalise the ${\cal Z}_3$ transformation
(see ref.\cite {fat})in $W_3$ to the case of $W_N$.
This transformation ${\cal Z}({\cal Z}^N=1)$ acts on the
weight vectors ($\mu , \nu $) $\equiv $ ($\sum _i n_i \Lambda ^i$, $\sum _i
m_i \Lambda ^i$) of $SU(N)_k$ and $SU(N)_{k+1}$ respectively as
\be
(n_1,n_2,.......,n_{N-1})~\stackrel {{\cal Z}} {\longrightarrow }
(\left [k~-~\sum_ i
{}~n_i\right ],n_1,.......n_{N-2}) \label {tran}
\ee
\be
(m_1,m_2,.......,m_{N-1})~\stackrel {{\cal Z}} {\longrightarrow }
(\left [k+1~-~\sum_ i
{}~m_i\right ],m_1,.......m_{N-2}) \label {trans}
\ee
This transformation leaves the conformal weight ($h _{\mu ,\nu }$)
invariant.

The conformal weights can also be rewritten as
$$
h_{\mu ,\nu } = h_{\mu }^{(k)} - h_{\nu }^{(k+1)} + {(\mu - \nu )^2 \over 2}
$$
where
$h_{\mu }^{(k)} = {\mu . (\mu + 2\rho) \over 2 (k+N)}$
is the conformal weight  of $SU(N)$ representation $\mu $ in the
$SU(N)_k$ WZW model.

The N copies of each primary field in $W_N$ models will be obvious if we
rewrite the conformal weight incorporating the level 1 field labelled
by $\eps$ as
follows:
\be
h_{\mu , \nu } = h_{\mu }^{(k)} + h_{\eps }^{(1)} -
h_{\nu }^{(k+1)} + {
(\mu - \nu )^2 - (\eps)^2 \over 2 }
\ee
where $\eps $ is such that $\eps . \psi \leq 1$
(i.e. $\eps$ is a singlet or any of the $(N-1)$ fundamental
representations of $SU(N)$)
and is given by
${\mu  - \nu } ~= ~ \eps $~~mod ~root~.

Two copies $\Phi _{(\mu_0,\nu_0)}$, $\Phi _{(\mu_1 ,\nu_1)}$ are related
by the ${\cal Z}$ transformation through a phase:
$$\Phi _{(\mu _0,\nu _0)}=
\exp \left ((N-1){\pi i\over 2} (\mu_1
- \nu_1+\eps _1).\eps _1\right )
\Phi _{(\mu _1,\nu_1)}$$
where ${\mu_i  - \nu_i } ~= ~ \eps _i$~~mod ~root~; $\eps _0$ is a
singlet; and hence $(\mu _0-\nu _0)=$ a root; and
$\mu _1 = {\cal Z}\mu _0$, $\nu _1 = {\cal Z}\nu _0$, $\eps _1={\cal
Z}\eps _0$.
The transformed weight vectors: ${\cal Z}\mu _0$ , ${\cal Z}\nu _0
$ are
given by eqns.(\ref {tran}) and (\ref {trans}) respectively and
${\cal Z}\eps _0$ is the corresponding transformed weight vector of
$SU(N)_1$.
Other copies are generated by transformation ${\cal Z}^2$,
${\cal Z}^3$, $\cdots$, ${\cal Z}^{N-1}$ on $\Phi _{(\mu _0,\nu _0)}$.

The fusion rules satisfied by these primary fields are
\be
\Phi_{(\mu _1,\nu _1)} \otimes \Phi_{(\mu _2,\nu _2)} ~~=~~
\sum _{\mu _3} \sum _{\nu _3} ~~\Phi _{(\mu
_3,\nu _3)}
\ee
where $\mu _3$ and $\nu _3$ are the representations allowed by the
fusion rules of two $SU(N)_k$ and $SU(N)_{k+1}$ Wess-Zumino models
respectively.

The braiding eigen values are given by
\be
\l_{(\mu_3,\nu _3)}[(\mu_1,\nu_1);(\mu _2,\nu _2)]~~=~~\Omega \exp (\pi i
h_{\mu_1,\nu_1}+h_{\mu_2,\nu_2}-h_{\mu_3,\nu_3})
\ee
where $\Omega~=(-1)^{f_1+f_2-f_3}$
where $f_i$ is an integer given in terms of the weight vectors by
$$
f_i={1\over 2}(\mu_i -\nu_i +\eps_i).
(\mu_i -\nu_i -\eps_i +2\rho )
$$
This braiding eigen value can be rewritten in terms of the $SU(N)$
WZW eigenvalues:
\be
\lambda_{(\mu_3,\nu _3)}[(\mu_1,\nu_1);(\mu _2,\nu _2)]~~=\lambda _{\mu
_3}^{(k)}(\mu _1,\mu _2)\left (\lambda _{\nu _3}^{(k+1)}(\nu _1,\nu _2)
\right )^{-1}
\lambda _{\eps _3}^{(1)}(\eps _1,\eps _2)
\ee
We expect that the duality matrix also factorises. Thus theorem 1
generalises to the $W_N$ model :

\noindent {\bf Theorem 3:}
The invariant for a link L, made up of $n$ knots, carrying
representations
$\Phi
_{(\mu_1,\nu_1)},......,\Phi _{(\mu_n,\nu_n)}$ respectively
is given in terms of
$SU(N)_k $, $SU(N)_{k+1}$ and $SU(N)_1$ invariants as follows:
\be
V_{(\mu _1,\nu _1),.......,(\mu _n,\nu_n)}[L]~~=~~V_{\mu_1,.......,\mu
_n}^{(k)}[L]~~
V_{\nu_1,.......,\nu_n}^{(k+1)}[\bar L]~~V_{\eps_1,.....,
\eps _n}^{(1)}[L]~~
\ee
where $\mu_i - \nu _i = \eps _i $mod~ root.

These results reflect the quantum group structure of superconformal and
$W_N$ models to be
$SU(2)_{q(k)}\otimes SU(2)_{q(2)}\otimes$ $SU(2)_{q^*(k+2)}$
and $SU(N)_{q(k)}\otimes SU(N)_{q(1)}\otimes SU(N)_{q^*(k+1)}
$ respectively.
\vskip1cm
\noindent{\bf 3.Concluding Remarks}

In this paper we have argued following earlier works
\cite{blok,moor} and generalising our methods of
computation from Chern-Simons theories \cite{wit,ours,ours1,
kaul} that factorisation of the braid and duality matrices
completely simplifies the corresponding knot/link invariants
for any rational conformal field theory
admitting a coset representation. The invariants from such a theory are
given in terms of Wess-Zumino theories based on the factors in the
coset.
In particular, minimal, superconformal
and $W_N$ model invariants are given by the product of invariants of
Wess-Zumino models as
given in Theorems 1,2 and 3.
While one can completely write down the  knot/link
invariants for this type of coset models we also find that
we do not get any more new invariants other than that given
by Chern-Simons theories for compact semi-simple Lie groups.

At this stage it is natural to ask the question: Is there a Chern-Simons
description of the invariants above obtained from conformal field
theories?
To answer this, let us focus our attention for definiteness on the
above
invariants obtained from $(G_k\otimes G_l)/ G_{k+l}$ theories.
These knot invariants can be given in terms of expectation values of
certain definite Wilson operators in three Chern-Simons theories based
on gauge groups G and couplings $k$,$l$ and $k+l.$
All possible Wilson link operators are given by the product of Wilson
link operators of these three theories.
For a link $L$ made up of $n$ component knots $K^{(1)}, K^{(2)},...,
K^{(n)}$, we place the representations ($R_1^{(i)}, R_2^{(i)},
S^{(i)}$) of the three gauge groups G respectively on the component knot
$K^{(i)}$. Then the Wilson link operator are:

$$W_{\{R_1\}~\{R_2\}~\{S\}}[L] =\prod_{i=1}^n
W_{R_1^{(i)}}^{(k)}[K^{(i)}]
\prod_{j=1}^n W_{R_2^{(j)}}^{(l)}[K^{(j)}]
\prod_{m=1}^n W_{S^{(m)}}^{(k+l)}[K^{(m)}]$$
$$=W_{\{R_1\}}^{(k)}[L] W_{\{R_2\}}^{(l)}[L]
W_{\{S\}}^{(k+l)}[L]$$
where the three factors are the link operators for the Chern-Simons
theories based on gauge group G with couplings respectively $k, l, k+l$
and set of representation $\{R_1\}, \{R_2\} and \{S\}$ living on the
component knots. Of all possible representation ($R_1^{(i)}, R_2^{(i)},
S^{(i)}$), only a subset can be associated with the primary fields
$\Phi_{(R_1^{(i)},R_2^{(i)},S_{}^{(i)})}$ of the conformal field theory
based on
$(G_k\otimes G_l)/G_{k+l}$. For this restricted but invariant
set of representations,
construct the following product of Wilson link operators:
\be
W_{\{R_1\}~\{1\}~\{1\}}[L]
W_{\{1\}~\{R_2\}~\{1\}}[L]
W_{\{1\}~\{1\}~\{S\}}[\bar L] =
W_{\{R_1\}}^{(k)}[L] W_{\{R_2\}}^{(l)}[L]W_{\{S\}}^{(k+l)}[\bar L]
\label {last}
\ee
where $\bar L$ is the mirror image of link $L$
 and ${1}$ reflects that all component knots carry the singlet
representation in the corresponding Chern-Simons theory. Then the invariant
associated with link $L$ obtained  from the
$(G_k\otimes G_l)/ G_{(k+l)}$ conformal field theory
with fields
$\phi _{(R_1^{(i)},R_2^{(i)},S_{}^{(i)})}$
placed on the component knots $K^{(i)}$ is given by the functional
average of the operator (\ref {last}) in the product Chern-Simons
theory:
\be
V_{(R_1^{(1)},R_2^{(1)},S^{(1)}),\cdots ,(R_1^{(n)},R_2^{(n)},S^{(n)})}[L]=
\langle~ W_{\{R_1\}}^{(k)}[L]W_{\{R_2\}}^{(l)}[L]W_{\{S\}}^{(k+l)}[\bar
L]\rangle
\ee

\vskip2cm

\flushleft{\bf Figure Captions:}\\
\begin{itemize}
\item[Fig.1] Three-ball $B_1$ with $S^2$ boundary
 represented by $\ket {\psi _0}$.
\item[Fig.2] Three-ball with $m$ half-twists in the central two parallel
strands represented by $\ket {\psi _m}$.
\item[Fig.3] Three-ball with $S^2$ boundary
 represented by $\ket {\psi ^{\prime}_0}$.
\item[Fig.4] Three-ball with $m$ halftwists in the side strands represented by
$\ket {\psi ^{\prime}_m}$.
\item[Fig.5]Hopf links drawn in two different ways--closure of a parallel
2-braid and an antiparallel 2-braid with two  half-twists.
\item[Fig.6]
Consistency condition for duality matrix factorisation.
\end{itemize}

\end{document}